\documentclass[aps,prl,twocolumn,superscriptaddress]{revtex4-1}

\usepackage{graphics}
\usepackage{graphicx}
\usepackage{mathtools}
\usepackage{verbatim}
\usepackage{amsmath}
\usepackage{appendix}
\usepackage{subfigure}
\usepackage{xcolor}
\newcommand{\angstrom}{\text{\normalfont\AA}}

\begin{document}

\title{Compact ultracold neutron source concept for low energy accelerator-driven neutron sources}

\author{Yun Chang Shin}
\email[]{corresponding author : yunshin@ibs.re.kr}
\affiliation{Center for Axion and Precision Physics, IBS, 193 Munji-Ro, Daejeon, South Korea, 34051}

\author{W. Michael Snow}
\author{David V. Baxter}
\author{Chen-Yu Liu}
\affiliation{Department of Physics, Indiana University, 727 E. Third St., Bloomington, IN, USA, 47405}
\affiliation{Center for Exploration of Energy and Matter 2401 Milo B. Samson Lane, Bloomington, IN, USA, 47408}

\author{Dongok Kim}
\author{Younggeun Kim}
\author{Yannis K. Semertzidis}
\affiliation{Center for Axion and Precision Physics, IBS, 193 Munji-Ro, Daejeon, South Korea, 34051}
\affiliation{Department of Physics, KAIST, 291 Daehak-Ro, Daejeon, South Korea, 34141}

\date{\today}

\begin{abstract}
The concept of a small-scale, pulsed-proton accelerator based compact ultracold neutron (UCN) source is presented. The essential idea of the compact UCN source is to enclose a volume of superfluid $^{4}\mathrm{He}$ converter with a supercold moderator in the vicinity of a low-radiation neutron production target from (p, n) reactions. The supercold moderator should possess an ability to produce cold neutron flux with a peak brightness near the single-phonon excitation band of the superfluid $^{4}\mathrm{He}$ converter, thereby augmenting the UCN production in the compact UCN source even with very low intensity of neutron brightness. The performance of the compact UCN source is studied in terms of the UCN production and thermal load in the UCN converter. With the proposed concept of the compact UCN source, a UCN production rate of $P_{\mathrm{UCN}}=80\mathrm{UCN}/\mathrm{cc}/\mathrm{sec}$ in the UCN converter could be obtained while maintaining thermal load of on the superfluid $^{4}\mathrm{He}$ and its container at a level of $22\mathrm{mW}$. This study shows that the compact UCN source can produce a high enough density of UCN at a small-scale, low-energy, pulsed-proton beam facility with reduced efforts on the cooling and radiation protection.

\end{abstract}

\maketitle

%%%%%%%%%%%%%%%% 
\section{\label{sec:intro}Introduction}
%%%%%%%%%%%%%%%% 
Neutrons with a kinetic energy of less than $250\mathrm{neV}$ are normally referred to as ultracold neutrons (UCNs). These ultra low-energy neutrons can be totally reflected from the surface of certain materials with high optical potential energy ($V_\mathrm{F}$) at any incident angle. This allows us to store UCNs in a material container for times which are virtually limited only by the neutron lifetime of about 880 seconds\cite{Patrignani_2016}. UCNs can also be highly polarized by passing them through magnetic fields, and can be trapped by strong magnetic field gradients. 

Due to these favorable properties, UCNs have been employed as a sensitive probe in physics experiments where high-precision measurements are required\cite{2014arXiv1412.5013A}. Various tests with UCNs have especially attempted to address unanswered questions in fundamental physics, astrophysics, and cosmology\cite{2009JPhG...36j4001N,2018arXiv181102340A}. These include the testing of fundamental theories such as searching for a neutron electric dipole moment (nEDM)\cite{Ito:2007fk,Lamoreaux:2009fk,Ahmed_2019,2020PRL..124.081803}, measuring the neutron lifetime\cite{Pichlmaier_2010,2014PhRvC..89e2501S,Serebrov_2018,2011RvMP...83.1173W} and neutron decay correlation coefficients\cite{2014JPhG...41k4007Y}, searching for axions and axion-like particles\cite{2007PhRvD..75g5006B,2009PhLB..680..423S,2015PhLB..745...58A,2003PhRvD..67j2002N,2011NatPh...7..468J}, searching for certain forms of dark matter\cite{2015IJMPA..3030048P}, and testing exotic long-range interactions \cite{Haddock_2018_plb,Haddock_2018,2020PhRvD.101l2002P}

The most common impediment in these experiments with UCNs is, however, a statistical uncertainty due to the limited number of UCNs available for counting from UCN sources. For any experiment with UCNs, the maximum available number of UCN is essential to make their measurements more meaningful with reduced statistical errors. For example, in the search for nEDM with UCNs, the statistical limit ($\sigma$) is totally limited by the number of UCNs ($N$) for the count as
\begin{equation}
\label{eq0}
\sigma({d}_{n})\approx\frac{\hbar}{2 \alpha\times E\times T\times\sqrt{N}},
\end{equation}
where the visibility ($\alpha$), the electric field  strength ($E$), and NMR coherent time ($T$). As seen in Eq. \ref{eq0}, to increase the sensitivity of nEDM experimental searches with UCNs, it is necessary to increase  the number of UCN available for the experiment from UCN sources. All other precision experiments with UCNs have the same demand on more UCNs for better counting statistics. 

Up to now, the brightest UCN sources operate in what is termed the ``superthermal'' mode in which neutrons do not come into thermal equilibrium with the medium of UCN conversion\cite{1975PhLA...53..133G}. In this case, cold neutrons are incident on the UCN converter medium held at a low temperature and lose their energy in one step by exciting the collective modes of the medium, conserving energy and momentum, and coming nearly to rest in the medium, thereby becoming UCNs. The upscattering of UCNs to higher energy states can be suppressed by lowering the temperature of the UCN converter medium, so that no thermal excitation is present\cite{Golub:1979fk}. Still the main limitation to increase  the UCN density in the UCN converter comes from the loss of UCNs by nuclear absorption in the medium. Superfluid $^{4}\rm{He}$\cite{springerlink:10.1007/BF01307673}, solid $\rm{D}_{2}$\cite{PhysRevB.62.R3581}, and solid $\rm{O}_{2}$\cite{2004nucl.th...6004L,2010arXiv1006.2970F} are the most commonly tested media for use as UCN converters thus far due to their zero or very low nuclear absorption cross-sections and the existence of elementary excitations (phonons, magnons, etc.) whose dispersion relation intersects with that of the free neutrons. 

Among them, solid $\rm{D}_{2}$ is known to have high UCN conversion rate (a factor of 30 larger than superfluid $^{4}\rm{He}$) from molecular rotational and phonon modes in solid state $\rm{D}_{2}$ molecules\cite{PhysRevB.62.R3581,Golub:1983fk,Adamczak:2011fk}. Especially, due to the large spin-relaxation energy of $\sim7.5\mathrm{meV}$, the kinetic energy of incident cold neutrons doesn't need to be very low.  In addition, solid $\rm{D}_{2}$ UCN source can be operated at a temperature of $\sim 5\mathrm{K}$ which is higher than the typical operating temperature of superfluid $^{4}\mathrm{He}$\cite{PhysRevB.62.R3581}. Therefore, the cryogenic requirement is less demanding for operating a solid $\rm{D}_{2}$ based UCN source. These two features make solid $\rm{D}_{2}$ as a practical UCN converter medium for UCN sources cited at reactor or spallation neutron facilities.
 
However, there are a couple of significant downsides to the solid $\rm{D}_{2}$ as an efficient UCN converter medium. One of the main drawback is the extremely short  lifetime of UCN  in the solid $\rm{D}_{2}$. After UCNs are produced in the solid $\rm{D}_{2}$, they  get lost within less than $100\mathrm{msec}$ through UCN upscattering or nuclear absorption\cite{2002PhRvL..89A2501M}. The other disadvantage is very short mean free path of UCNs in the solid $\rm{D}_{2}$ which is about $2\sim3\mathrm{cm}$\cite{2018EPJA...54..148A}. It causes UCN extraction efficiency to be dropped as the thickness of solid $\rm{D}_{2}$ increases. As a result, the solid $\rm{D}_{2}$ converters have a certain restraint in terms of thickness of the converter volume.

Unlike solid $\rm{D}_{2}$, the mean free path of neutrons with a wavelength of $8.9\mathrm{\AA}$ in superfluid $^{4}\rm{He}$ can be as long as $17\mathrm{m}$\cite{1955PhRv...97..855S}. In addition, the isotopically pure $^{4}\rm{He}$ can ideally have zero nuclear absorption cross section. Therefore, a longer length of the UCN converter volume with superfluid $^{4}\rm{He}$ can be made to increase UCN production with superfluid $^{4}\rm{He}$ UCN converter. For superfluid $^{4}\rm{He}$ as a UCN converter medium, the observation showed that only a certain energy range of cold neutrons would play an important role in UCN production due to the well-known phonon-roton dispersion curve in superfluid $^{4}\rm{He}$\cite{Gibbs1996thesis}. 

The production mechanism of UCNs in superfluid $^{4}\rm{He}$ can be decomposed into two distinct processes: single-phonon excitation and multi-phonon excitation\cite{Gibbs1996thesis,2002PhLA..301..462K}. The multi-phonon process involves a broad range of neutron dynamic structure factor $\mathrm{S}(\mathrm{Q}, \hbar\omega)$ in the energy-momentum space. On the other hand, the neutron dynamic structure factor for single-phonon excitation is very sharply defined, and forms nearly a delta function in the dispersion curve\cite{Gibbs1996thesis,Schmidt-Wellenburg:1uq}. Owing to this characteristic of superfluid $^{4}\rm{He}$ as a UCN convertor, only a very narrow energy range of cold neutrons near $1\mathrm{meV}$ (corresponding to a neutron wavelength of $8.9\mathrm{\AA}$) can be downscattered into the UCN energy regime through the single-phonon excitation. 

However, the great emphasis would be placed on the difference between two excitation processes in the UCN conversion probability: the single-phonon excitation  process is more than one order of magnitude more probable than the multi-phonon process at the $T \to 0$ limit\cite{Schmidt-Wellenburg:1uq}. On account of above observation, it becomes obvious that the main key to enhance the UCN production in a superfluid $^{4}\mathrm{He}$ converter is to increase the differential flux of incoming neutrons with their kinetic energy near $1\mathrm{meV}$  so that more cold neutrons can be downscattered to UCNs through the single-phonon excitation mode than multi-phonon mode.

So far, the superthermal UCN source has been realized only at large neutron facilities, either spallation or reactor-based neutron source, where high flux of neutron is available. However, superfluid $^{4}\mathrm{He}$ UCN sources cited at these facilities have yet to meet this condition, more UCN production through the single-phonon excitation than multi-phonon excitations, properly due to their limited moderation power to produce cold neutron spectrum\cite{Masuda:2012uq,2019arXiv191208073S}. 

Liquid $\mathrm{H}_{2}$ and liquid $\mathrm{D}_{2}$ are the most common cold moderator media available in the high power neutron facilities. Typical cold neutron spectra from these moderators show their peak flux near a few $\mathrm{meV}$, but drop quickly in the region of $1\mathrm{meV}$ where the single-phonon excitation band of superfluid $^{4}\mathrm{He}$ exists. Therefore, the efficiency of UCN production in the $^{4}\mathrm{He}$ would still be rather poor. For this reason, superfluid $^{4}\mathrm{He}$ UCN sources cited at these facilities have a tendency to increase overall neutron flux to get more UCNs by increasing proton beam power or locating the neutron source close to reactor source.

However, neutrons away from this energy introduce an additional thermal load on the moderator vessel and medium through neutron capture reactions. If one is too close to the neutron source, the heat load from fast neutrons and gammas emitted from the reactor core or spallation target becomes prohibitive.  There are other superfluid $^{4}\mathrm{He}$ UCN sources located at the end of cold neutron beam lines accepting only neutrons with an energy of $1\mathrm{meV}$. In this case, these sources can handle reduced heat loads but still have to compromise with low UCN density due to the limited intensity of cold neutron flux after the filtering.

\begin{figure}[htbp]
\begin{center}
\resizebox{0.45\textwidth}{!}{
  \includegraphics{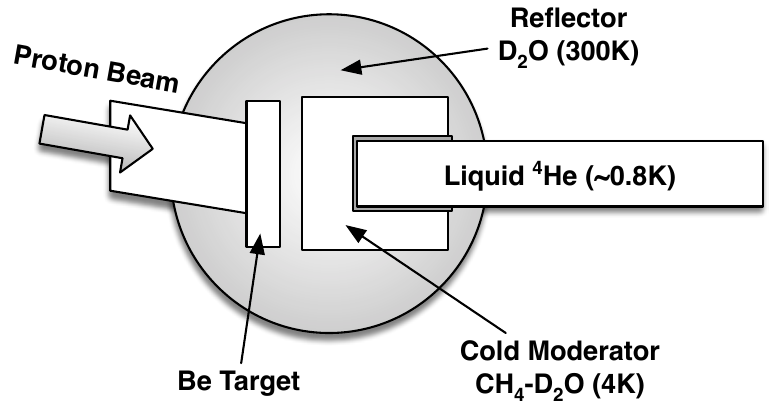}
}
\caption{Schematic of the compact UCN source. A proton beam is incident on a Be target. A heavy water reflector surrounds the target and cold moderator. The cold moderator surrounds the UCN production volume filled with superfluid $^{4}\mathrm{He}$.}
\label{fig1}
\end{center}
\end{figure}

In this paper, we propose UCN production in a compact UCN source. The compact UCN source utilizes superfluid $^{4}\rm{He}$ as the UCN converter medium. But the primary difference of the compact UCN source from other UCN sources is a benefit from colder neutron spectrum. To maximize the UCN production through the single-phonon excitation mode in the superfluid $^{4}\rm{He}$, cold neutrons with a peak flux of $\sim1\mathrm{meV}$ are fed from a supercold moderator held at $\sim4\mathrm{K}$ temperature. Unlike existing cold moderator media, a  medium of the supercold moderator proposed here can cause the peak energy of the cold neutron flux approaching toward $1\mathrm{meV}$. This will enhance the UCN production rate with minimal heat load on the UCN converter.  

As shown later, the UCN density will be improved at least one order of magnitude more than existing superfluid $^{4}\rm{He}$ UCN sources with the supercold moderator medium in our estimation. The much less intense radiation coming from the production target of the compact UCN source presents a sufficiently low heat load on the cold moderator volume to envision the operation of the supercold moderator. Therefore, the compact UCN source proposed here can produce a high density of UCNs to be useful enough for experiments without imposing prohibitive cryogenic or radiation safety requirements. 

%%%%%%%%%%%%%%%% 
\section{\label{sec:source}Compact UCN Source}
%%%%%%%%%%%%%%%% 
The compact UCN source in this paper represents a small scale, accelerator-based UCN source.  A conceptual design of the compact UCN source is shown in Fig. \ref{fig1}.  In the compact UCN source, a small-scale low-energy proton accelerator such as a linear accelerator (LINAC) or a cyclotron generates proton beam with energy of 13MeV and current of 25mA. The pulse width and frequency of the beam are set to be 0.6 msec and 20Hz, respectively. Recently, there are a number of projects being considered to improve the power levels of proton beams to 50 kW or more in the energy range of $2\sim50\mathrm{MeV}$ for compact accelerator-driven neutron sources (CANS) around the world \cite{R_cker_2016,Gutberlet2018physicaB}.  The compact UCN source can also be advanced by adapting the technical improvement from CANS\cite{Carpenter_2019}.

The proton beam is incident on a target to release neutrons with $\sim$MeV energy from $(\mathrm{p},\mathrm{n})$ reactions\cite{2007chris}. These neutrons produced from the $(\mathrm{p},\mathrm{n})$ reactions are less energetic than those from spallation or reactor sources, thereby remaining below the threshold of many secondary nuclear reactions. Metallic Be is chosen as the target material due to its high neutron yield, high melting point, good mechanical strength, and simplicity of handling. The thickness of the target is $1.1\mathrm{mm}$ which is chosen to be less than the range of the protons (1.3mm for a 13MeV proton beam in Be) in order to avoid the problems of hydrogen embrittlement of the Be. 

The neutron yields have been empirically measured for proton energy up to $23\mathrm{MeV}$\cite{1977NucIM.143..331L}. An empirical formula for the total neutron yield, ${Y}_{n}$, as a function of the proton energy, ${E}_{p}$ in $\mathrm{MeV}$, is
\begin{equation}
Y_{n}({E}_{p})=3.42\times10^{8}({E}_{p}-1.87)^{2.05} (\mathrm{n}/\mu\mathrm{C})
\end{equation}
This gives total number of neutron produced from the Be target as $\sim1.43 \times {10^{13}}{\rm{n}}$\cite{2007chris}.  

A volume of neutron reflector enclosing the Be target, cold moderator and the UCN converter is filled with $\rm{D_{2}O}$ at room temperature as it preserves and feeds more neutrons to the cold moderator mainly due to the very small neutron absorption cross-section of deuterium. The UCN converter volume is in the shape of a cylindrical tube in the horizontal direction. The tube made with a $2\mathrm{mm}$-thick high-purity aluminum has a total length of $\sim1\mathrm{m}$ and an inner diameter of $80\mathrm{mm}$. The medium for the UCN converter is an isotopically-pure superfluid $^{4}\rm{He}$ maintained at a temperature of $\sim0.8\mathrm{K}$. The pure $^{4}\rm{He}$ can be prepared with superleak method to maintain the level of $^{3}\rm{He}$ below 1 ppb (part per billion). The cold moderator volume has a shape of cylinder so that it symmetrically envelops the UCN converter volume to deliver maximum cold neutron flux into the UCN converter. A medium of the cold moderator is contained in a $2\mathrm{mm}$-thick high-purity aluminum housing, and the temperature of the cold moderator is maintained at 4K.  

Given that the activation from the reaction in the neutron production target is very low, the compact UCN source is capable to accommodate certain materials for advanced cold neutron moderation, which may not survive intact in environments with higher levels of radiation, to maximize the UCN production. Low Energy Neutron Source (LENS) at Indiana University, one of the first compact accelerator-driven neutron sources, has been successfully operating a solid methane cold moderator, which is the brightest slow neutron moderator medium known thus far, well into the solid phase at $4\mathrm{K}\sim20\mathrm{K}$\cite{2007chris,2010NIMPA.620..375S}. LENS can operate the solid methane moderator in spite of the well-known fact that radiation damage in solid methane can cause the moderator to explode\cite{1990cns..work....5C}. The radiation field produced in LENS is well below the threshold of the catastrophic recombination process of radiation defects which occurs in solid methane. 

By adapting the concept of cold neutron moderation from the compact neutron source, two different media are selected for the cold neutron moderator of the compact UCN source in this study. One is solid methane at $4\mathrm{K}$, and  the other is methane clathrate hydrate for reasons we will explain below. A variety of hydrogenous materials have been used or tested as a medium for cold moderators despite of a relatively large absorption cross section from hydrogen atoms. A new neutron-cooling mechanism using inelastic magnetic scattering in paramagnetic systems has also been recently proposed  for realizing intense sources of very cold neutrons whose energy range is loosely defined as sub-$\mathrm{meV}$ regime\cite{2016PhRvC..93c5503Z}.

The moderating performance of some hydrogenous materials was also evaluated in terms of effective neutron temperatures as a function of cold moderator temperature\cite{Inoue_1982}.  Among many of hydrogenous materials, solid methane is known as the best material for the cold neutron moderation due to its high-density of proton. Methane in solid state also possesses a large fraction of rotational degree of freedom that is effective to shift the neutron energy further down than any other elementary excitations so far\cite{1998JChPh.109.3161G}. 

\begin{figure}[htbp]
\begin{center}
\resizebox{0.35\textwidth}{!}{
  \includegraphics[width=6cm]{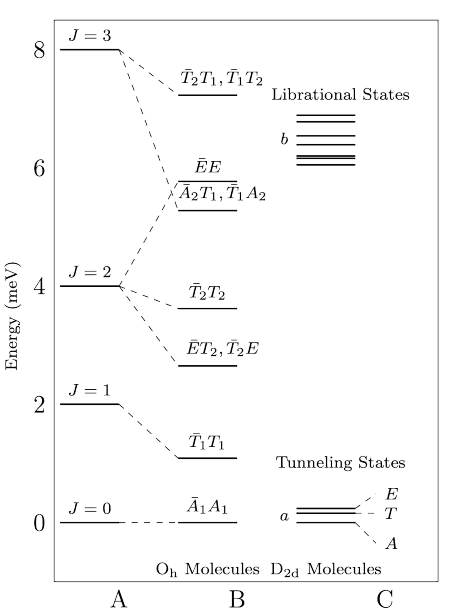}
}
\caption{The energy levels of two different spin states of $\mathrm{CH}_{4}$ molecules: (a) $O_{h}$ are the orientationally disordered molecules that cause free rotation of methane molecule; (b) $D_{2d}$ are the orientationally ordered molecules that cause the hindered rotation of methane molecule. Each state is related to a representation $A$, $T$ and $E$ corresponding to the total nuclear spin $I=2$(ortho), $I=1$(meta) and $I=0$(para), respectively.}
\label{fig10}
\end{center}
\end{figure}

Solid methane in phase I ($20\mathrm{K}\sim90\mathrm{K}$) shows almost free rotation in their molecular site. However, as they undergo lower temperature phase (phase II: $4\mathrm{K}\sim20\mathrm{K}$), only 1/4 of the molecules remain freely rotating. The rest of molecules undergo librations and tunneling motions between equivalent librational ground states.  This can be understood as disordered ($O_{h}$ symmetry) and ordered ($D_{2d}$ symmetry) states respect to molecular orientations\cite{1980JChPh..73.3442O}.  The energy levels of tunneling states in $D_{2d}$ symmetry are 0, 0.16, and $0.24\mathrm{meV}$, respectively. At the same time, the energies of the transitions between lower energy levels in $O_{h}$ symmetry are $1.09\mathrm{meV}$ for the $J=0\leftrightarrow J =1$ transition, $1.56\mathrm{meV}$ for the $J=1\leftrightarrow J=2$ transition, and $2.65\mathrm{meV}$ for the $J =0\leftrightarrow J =2$ transition, respectively.  Fig. \ref{fig10} shows the energy levels of each state.

While the energy level spacing among the tunneling states in $D_{2d}$ symmetry site appear to be rather too small for the effective down-shifting of neutron energy close to $\sim1\mathrm{meV}$, the level-spaced energy between ground and first excited states in the $O_{h}$ symmetry site would provides the right amount of energy shift for scattered neutrons, and makes a neutron spectrum to approach quickly toward $1\mathrm{meV}$\cite{2010NIMPA.620..375S,2010NIMPA.620..382S}. This makes solid methane at 4K as an effective medium for the cold neutron moderation.

On top of that, if there is any material possessing a larger fraction of methane molecules with the free rotational mode than normal solid methane in low temperature, it would be an even better material for the cold moderator to obtain a neutron flux rich in the very low energy regime especially near $1\mathrm{meV}$, and hence would be even more effective than solid methane when it couples to the superfluid $^{4}\mathrm{He}$ UCN converter.

With inferring the effectiveness of the free rotational mode in cold neutron moderation from solid methane, methane clathrate hydrate at 4K would be a strong candidate for the supercold moderator that is required for the compact UCN source due to the unique quantum rotations inside the clathrate cage. Methane clathrate hydrate ($\mathrm{CH}_{4}-\mathrm{D}_{2}\mathrm{O}$) is a nonstoichiometric inclusion compound of methane within the host framework of heavy ice ($\mathrm{D}_{2}\mathrm{O}$) cages in a unit cell\cite{2004PhyB..350E.395K}. These cages allow for nearly free rotation of methane molecules even at low temperatures. The molecular motion of methane is quantitatively described as a single-particle quantum rotor in a weak orientational potential, and thereby showing merely free rotational motion.  For $\mathrm{D}_{2}\mathrm{O}$ ice cages, they still preserve more neutrons from their low nuclear absorption cross section and coherent scattering nature even though they possesses a lack of low-energy excitation modes\cite{utsuro1981}. Therefore, the methane clathrate hydrate shows a combined attribute of high density of free rotational mode of methane molecules and coherent scattering propoerty of $\mathrm{D}_{2}\mathrm{O}$ ice, and provides a very effective down-scattering channel for a wide energy range of incoming neutrons.

\begin{figure}
\centering
\subfigure[Solid methane at 4K] 
{
    \label{fig:11:a}
    \includegraphics[width=7.5cm]{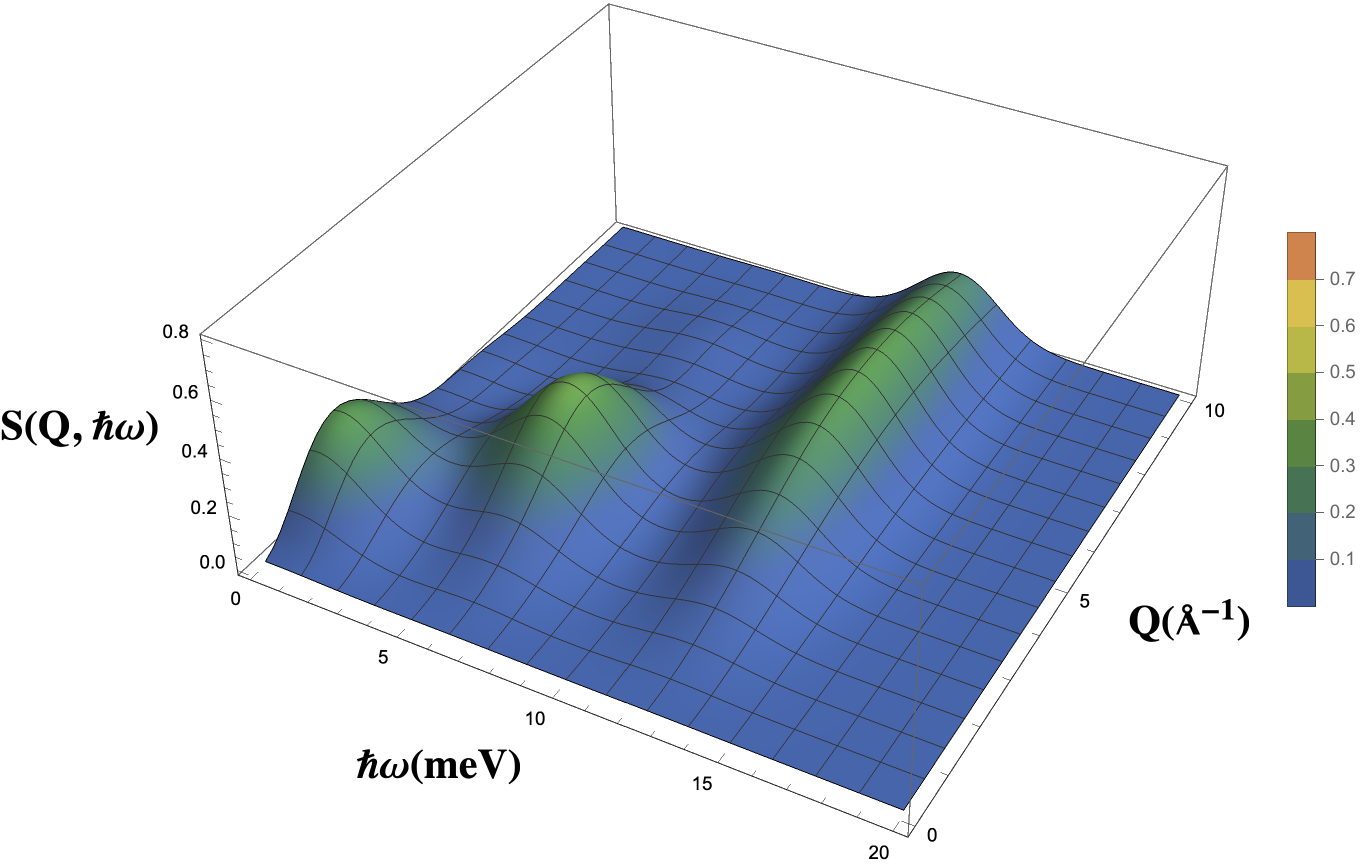}
} 
\hspace{1cm}
\subfigure[ ``Free rotational mode'' of single methane molecule in methane clathrate hydrate at 4K]
{
    \label{fig:11:b}
    \includegraphics[width=7.5cm]{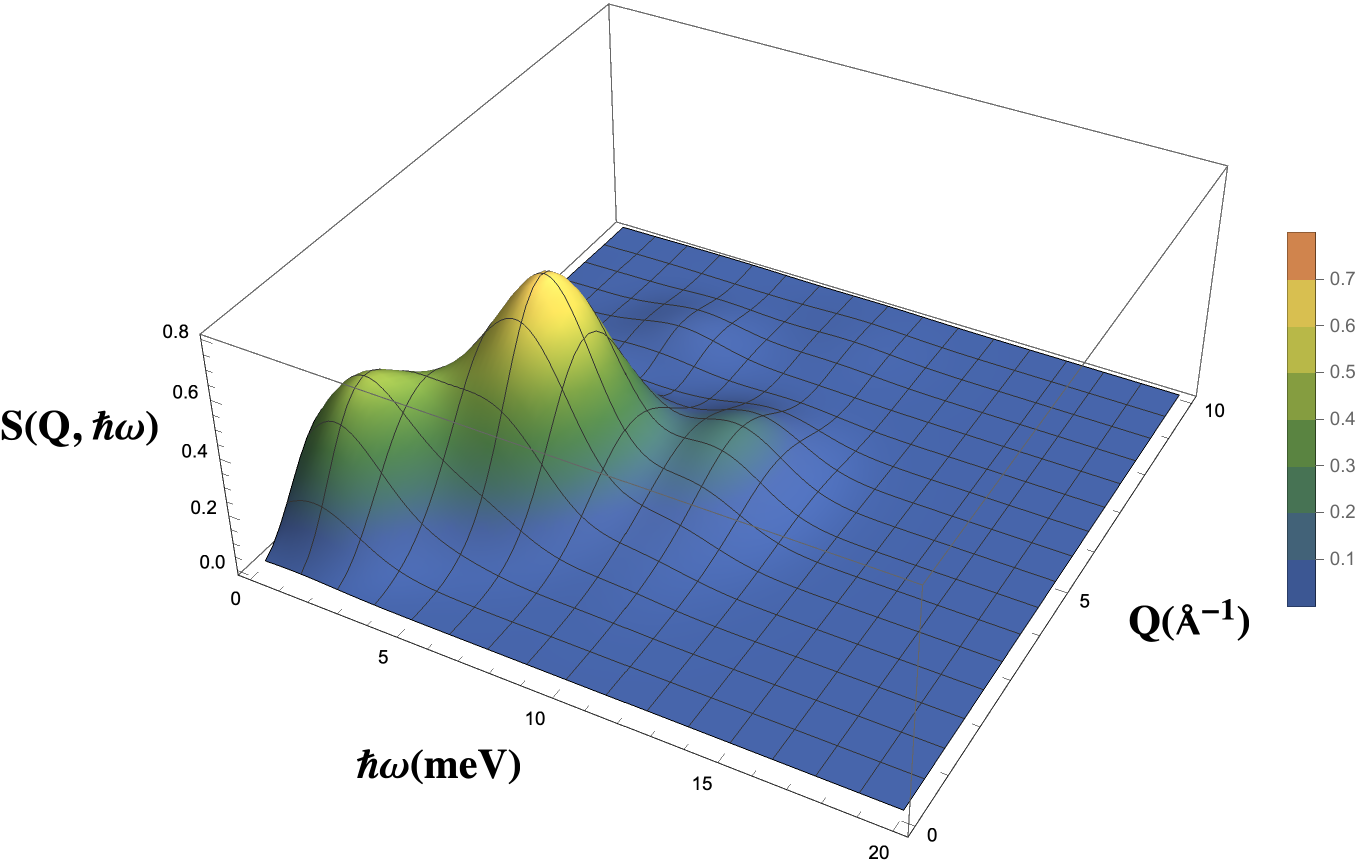}
} 
\caption
{Dynamic structure factor $\mathrm{S}(\mathrm{Q},\hbar\omega)$ of solid methane and methane clathrate hydrate at a temperature of 4 K. (a) Three peaks in solid methane correspond to (i) orientational tunnelings in $D_{2d}$ symmetry and low energy excitations in $O_{h}$ symmetry in low ($\mathrm{Q},\hbar\omega$) region, (ii) rotational librations in $D_{2d}$ and high energy excitations in $O_{h}$  in mid ($\mathrm{Q},\hbar\omega$) region, and  (iii) the multi-phonon excitation in high ($\mathrm{Q},\hbar\omega$) region, respectively. (b) Only the energy excitations in $O_{h}$ symmetry of methane molecules are shown for the $\mathrm{S}(\mathrm{Q},\hbar\omega)$ of methane clathrate hydrate.}
\label{fig:11}
\end{figure}

The neutron dynamic structure factor $\mathrm{S}(\mathrm{Q}, \hbar\omega)$ of solid methane in this temperature regime ($4\mathrm{K}\sim20\mathrm{K}$) was estimated and tested for the LENS\cite{2010NIMPA.620..375S,2010NIMPA.620..382S}. The neutron energy spectra estimated from a MCNPX\cite{2011MCNPX} simulation using this $\mathrm{S}(\mathrm{Q}, \hbar\omega)$ showed perfect agreement with those obtained from one of LENS neutron beamlines for moderator temperatures of 20K and 4K. This $\mathrm{S}(\mathrm{Q}, \hbar\omega)$ was also independently tested in a course of simulations conducted for a proposed UCN source based on an extracted thermal neutron beam\cite{Lychagin:2015aa}. To estimate the cold neutron flux from the methane clathrate, a scattering kernel carrying only the ``free rotations'' mode of methane molecules at $4\mathrm{K}$ is blended with the other kernel describing heavy water ($\mathrm{D}_{2}\mathrm{O}$)\cite{2010NIMPA.620..382S}. This blended scattering kernel shows almost identical characteristics features compared with the measured $\mathrm{S}(\mathrm{Q}, \hbar\omega)$ of methane clathrate hydrate ($\mathrm{CH}_{4}-\mathrm{D}_{2}\mathrm{O}$)\cite{2004PhyB..350E.395K,2006PhyB..385..202K}, thereby is still accurate enough for use in this study. Fig. \ref{fig:11} shows the dynamic structure factor $\mathrm{S}(\mathrm{Q},\hbar\omega)$ of solid methane and the ``free rotational mode" of methane molecules in methane clathrate hydrate at a temperature of 4K, respectively. 

\begin{figure}
\centering
\subfigure[Cold neutron flux in UCN volume]
{
    \label{fig6a}
    \includegraphics[width=7.5cm]{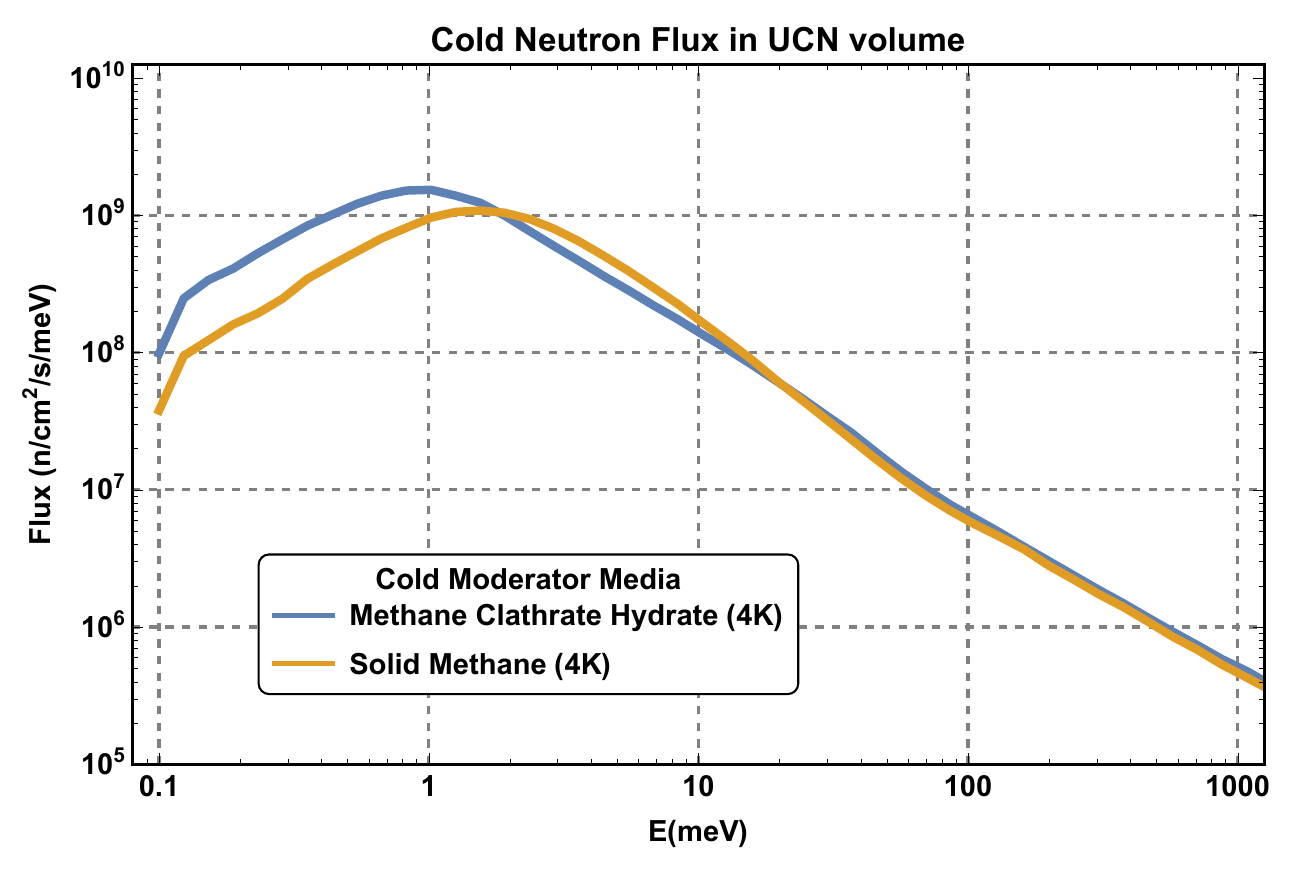}
} 
\hspace{1cm}
\subfigure[Cold neutron flux with $S(\lambda)$ of superfluid $^{4}\mathrm{He}$ ]
{
    \label{fig6b}
    \includegraphics[width=8.25cm]{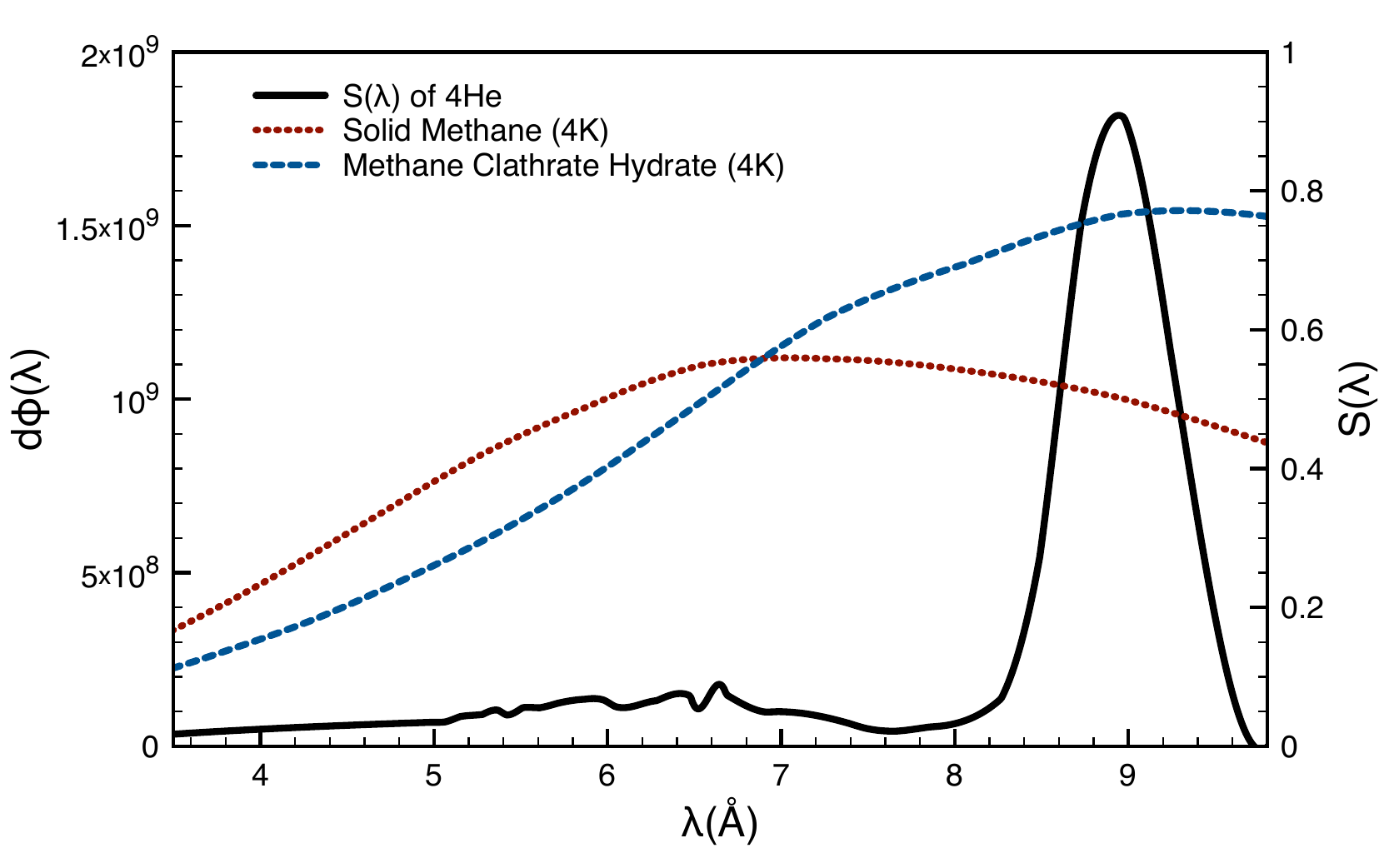}
} 
\caption
{(a) Cold neutron flux in the UCN converter volume from cold moderator media as function of neutron energy in $\mathrm{meV}$ (b) Cold neutron flux and $S(\lambda)$ of superfluid $^{4}\mathrm{He}$ as a function of the neutron wavelength in \AA.}
\label{fig6}
\end{figure}

The expected cold neutron flux in the UCN converter volume from these two moderator media is estimated by means of MCNPX. Fig. \ref{fig6a} shows the neutron spectra accumulated in the UCN converter volume from solid methane and methane hydrate clathrate respectively. The solid methane cold moderator produces a peak neutron flux of $1.12\times10^{9}\mathrm{n}/\mathrm{cm}^{2}/\mathrm{s}/\mathrm{meV}$ at an energy of $1.55 \mathrm{meV}$. The neutron flux at the energy of $1.0\mathrm{meV}$ is $9.47\times10^{8}\mathrm{n}/\mathrm{cm}^{2}/\mathrm{s}/\mathrm{meV}$. At the same time, methane hydrate clathrate cold moderator produces a peak neutron flux of $1.54\times10^{9}\mathrm{n}/\mathrm{cm}^{2}/\mathrm{s}/\mathrm{meV}$ at an energy of  $1.02\mathrm{meV}$ which is ideally close to the single phonon excitation band of superfluid $^{4}\mathrm{He}$ at a wavelength of $8.9\mathrm{\AA}$. 

%For comparison purposes, the cold neutron spectrum from liquid ortho-$\mathrm{D}_{2}$ cold moderator, which is one of the most common cold moderator media for current UCN sources, is also shown. The differential neutron flux at  $1\mathrm{meV}$ from  liquid ortho-$\mathrm{D}_{2}$ cold moderator is $3\times10^{7}\mathrm{n}/\mathrm{cm}^{2}/\mathrm{s}/\mathrm{meV}$ which is about two orders of magnitude lower than one from methane clathrate hydrate moderator. 

Fig. \ref{fig6b} shows the neutron spectra as a function of neutron wavelength from  solid methane and methane hydrate clathrate with the scattering function, $\mathrm{S}(\lambda)$, of superfluid $^{4}\mathrm{He}$. The numerical data of $\mathrm{S}(\lambda)$ at $0.5\mathrm{K}$ in SVP (saturated vapor pressure) was sourced from the other literature\cite{Schmidt-Wellenburg:1uq}. As shown, the peak wavelength of the neutron spectrum from methane clathrate hydrate cold moderator is perfectly matched with the wavelength of single-phonon mode where superfluid $^{4}\mathrm{He}$ has the highest intensity of $\mathrm{S}(\lambda)$.  

%%%%%%%%%%
\section{\label{sec:ucnproduction}The UCN production in the Compact UCN Source}
%%%%%%%%%%

The UCN production rate in the compact UCN source is estimated with the cold neutron flux from each moderator. The UCN production volume made of an $2\mathrm{mm}$-thick aluminum has a neutron optical potential of ${V_{\rm{F}}^{\rm{Al}} = 54{\rm{neV}}}$. If a beryllium(Be) coating on the inner surface of the aluminum tube is assumed, the optical potential of the UCN storage volume will increase to the Fermi potential of Be of $V_{\rm{F}}^{\rm{Be}}=252\mathrm{neV}$\cite{2016PhRvC..93b5501L}. This will not only enhance the UCN density significantly but will also extend the UCN storage time. 

The geometry and position of the UCN production volume itself are fixed. The total volume of the liquid $^{4}\mathrm{He}$ container is about $5320\mathrm{cc}$. The area surrounded by the cold moderator is $27\mathrm{cm}$ long from the bottom end. Therefore, the volume involved in the UCN production is $1350\mathrm{cc}$. The rest of the tube volume is considered as the UCN storage volume despite the absence of a geometrical change. 

UCN production in superfluid $^{4}\mathrm{He}$ is based on the energy-momentum relationship of the superfluid $^{4}\mathrm{He}$. From the coherent inelastic scattering of the incident cold neutrons with energy $E$, and momentum $k$ down to energy $E'$, momentum $k'$, the production rate per unit volume is determined from the literature \cite{Schmidt-Wellenburg:1uq,2002PhLA..301..462K}
\begin{equation}
\label{eq1}
{P_{\mathrm{UCN}}}({V_{\mathrm{c}}}) = \int_o^\infty {dE\int_0^{{V_{\mathrm{c}}}} {N\frac{{d\phi }}{{dE}}} } \frac{{d\sigma }}{{dE'}}(E \to E')dE',
\end{equation}
where $d\phi/dE$ is the differential incident flux of the cold neutron, $N$ is the number density of $^{4}\mathrm{He}$ , and $d\sigma /dE'$ is the differential cross-section for inelastic neutron scattering. The critical potential energy $V_{\mathrm{c}}$ of the UCN is given by the relative optical potential of the UCN container from the Fermi potential of superfluid $^{4}\mathrm{He}$ as $V_{\mathrm{c}}=V_F^{\mathrm{con}}-V_{F}^{^{4}\mathrm{He}}$. The differential cross-section is given by
\begin{equation}
\label{eq2}
\frac{{d\sigma }}{{dE}} = 4\pi {b^2}\frac{{{k^\prime }}}{k}\mathrm{S}(\mathrm{Q},\hbar \omega ),
\end{equation}
where $b$ is the bound neutron scattering length of $^{4}\mathrm{He}$, $\hbar\omega=E-E'$ is the energy transfer, $Q=k-k'$ is the momentum transfer, and $\mathrm{S}(\mathrm{Q},\hbar\omega)$ is the dynamic structure function evaluated in the energy-momentum space for free neutrons. For superfluid $^{4}\mathrm{He}$, Eq. \ref{eq1} can be expressed as 
\begin{equation}
\label{eq3}
{P_{\mathrm{UCN}}}({V_{\mathrm{c}}}) = N\sigma {V_{\mathrm{c}}}\frac{{{k_c}}}{{3\pi }}\int_0^\infty {\frac{{d\phi }}{{d\lambda }}} s(\lambda )\lambda d\lambda,
\end{equation}
where $\sigma$ is the bound neutron scattering cross section of $^{4}\mathrm{He}$. The scattering function, $s(\lambda)$, of superfluid $^{4}\mathrm{He}$ is defined as a function of the incident neutron wavelength $\lambda$,
\begin{equation}
\label{eq4}
s(\lambda ) = \hbar \int {\mathrm{S}(\mathrm{Q},\hbar \omega )\delta (\hbar \omega - {\hbar ^2}{k^2}/2{m})d\omega },
\end{equation}
where $Q=k=2\pi/\lambda$ with the assumption of $k'\ll k$. The scattering function can be divided into two parts, single-phonon and multi-phonon parts, as $s(\lambda ) = {s_{\rm{single}}}(\lambda ) + {s_{{\rm{multi}}}}(\lambda )$.

The single-phonon UCN production rate found in the literature\cite{2002PhLA..301..462K,Schmidt-Wellenburg:1uq} gives,
\begin{equation}
\label{eq5}
{P_{\rm{single}}}({V_{\mathrm{c}}}) = N\sigma {\left( {\frac{{{V_{\mathrm{c}}}}}{{{E^*}}}} \right)^{3/2}}\frac{{{\lambda ^*}}}{3}\beta {S^*}{\left. {\frac{{d\phi }}{{d\lambda }}} \right|_{{\lambda ^*}}},
\end{equation}
where $\lambda^{*} = 2\pi /q^{*}$ is the neutron wavelength at the intersection of the dispersion curves of the free neutron and the helium (${q^{*}} = 0.706{{\angstrom}}^{-1}$ for SVP), and ${S^*} = \hbar \int_{{\rm{peak}}} {\mathrm{S}(\mathrm{Q},\hbar\omega )} d\omega $ denotes the intensity due to single-phonon emission. In addition, $\beta=\frac{v_{n}^{*}}{v_{n}^{*}-c_{n}^{*}}$ is the Jacobian factor.

For the Be coated UCN converter, the critical potential energy $V_{\mathrm{c}}$ is defined as $V_{c}=V_F^{\mathrm{Be}}-V_{F}^{^{4}\mathrm{He}}=232\mathrm{neV}$. This yields the single-phonon production rate as
\begin{equation}
\label{eq6}
{P_{\rm{single}}}({V_{\mathrm{c}}}) = 4.93097 \times {10^{ - 8}}\frac{{\rm{A}}}{{{\rm{cm}}}}{\left. {\frac{{d\phi }}{{d\lambda }}} \right|_{{\lambda ^*}}},
\end{equation}
where ${\frac{{d\phi }}{{d\lambda }}}$ is the differential cold neutron flux near $\lambda=8.9\mathrm{\AA}$.

\begin{figure}[htbp]
\begin{center}
\resizebox{0.4\textwidth}{!}{
  \includegraphics{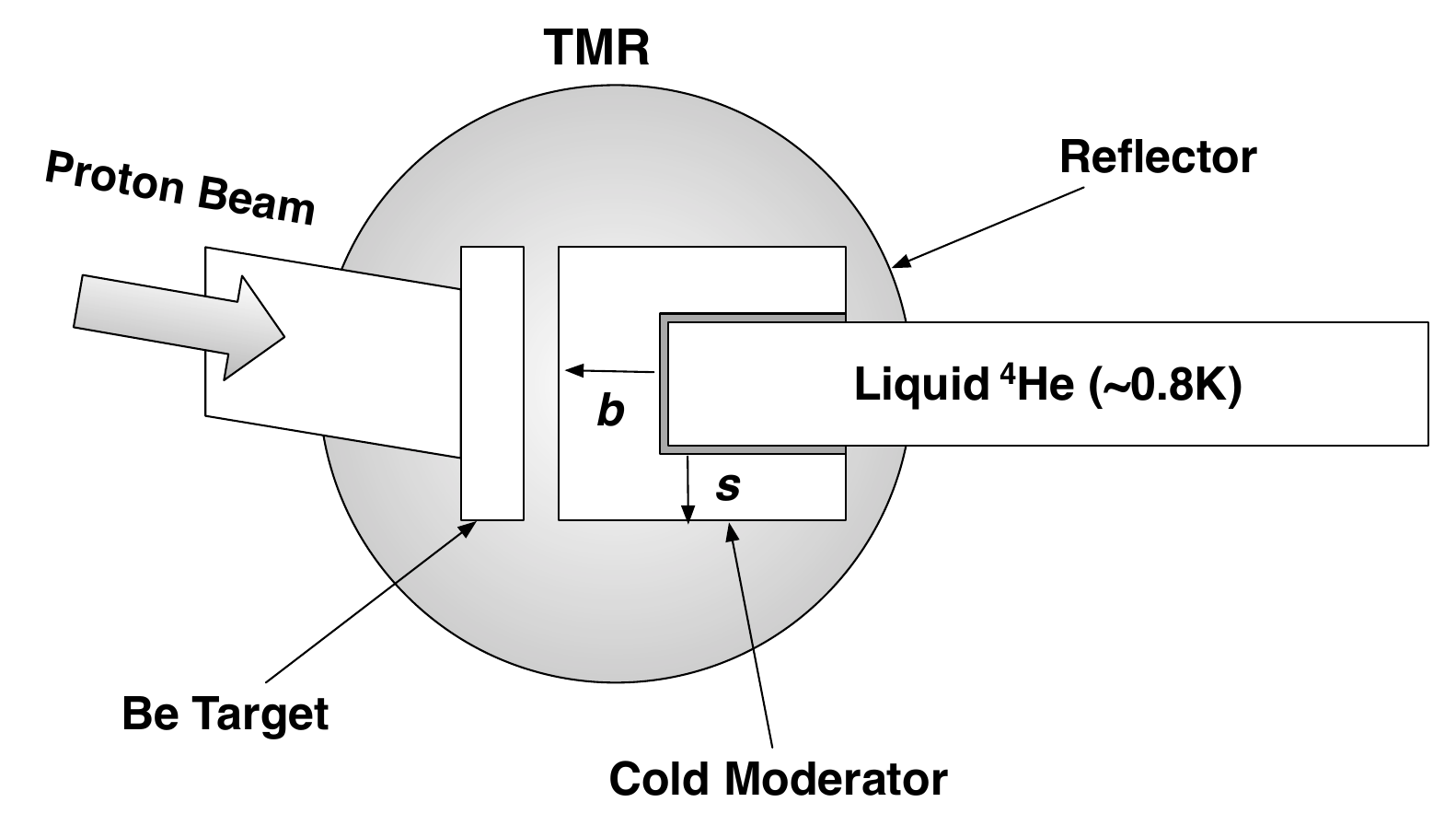}
}
\caption{Dimensions of the cold moderator volume. Side thickness $s$ and bottom thickness $b$ are varied to determine maximum UCN production rate.}
\label{fig2}
\end{center}
\end{figure}

The UCN production rate is estimated as a function of the cold moderator geometry while varying the bottom thickness $b$ and side thickness $s$ of the cold moderator volume as shown in Fig. \ref{fig2}. From Eq. \ref{eq3}, the UCN production rate is defined by the integral flux of the incident cold neutron. The proposed concept enables the feeding of cold neutrons from a broad angular distribution to the UCN production volume. In previous study of a cold neutron moderator for LENS, the optimal thickness of a solid methane moderator was about $2\mathrm{cm}$\cite{2010NIMPA.620..375S}.  In this study, the side thickness $s$ is varied between $0.5\mathrm{cm}$ to $2.2\mathrm{cm}$ while the bottom thickness $b$ is fixed at either 1cm or 2cm to determine the optimal thickness for maximum UCN production, 

\begin{figure}
\centering
\subfigure[Solid Methane (4K)]
{
    \label{fig3a}
    \includegraphics[width=7.5cm]{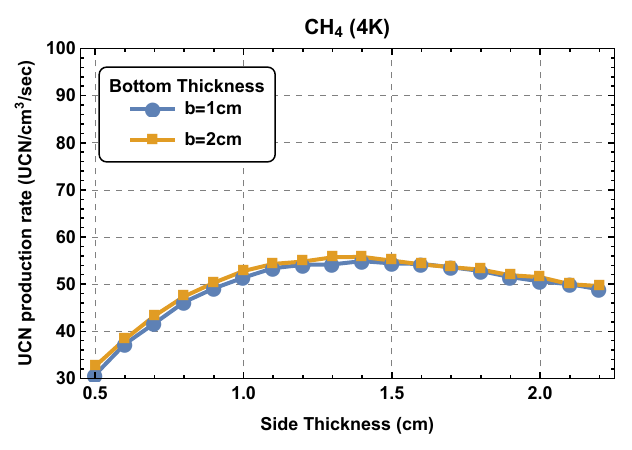}
} 
\hspace{1cm}
\subfigure[Methane Clathrate Hydrate (4K)] 
{
    \label{fig3b}
    \includegraphics[width=7.5cm]{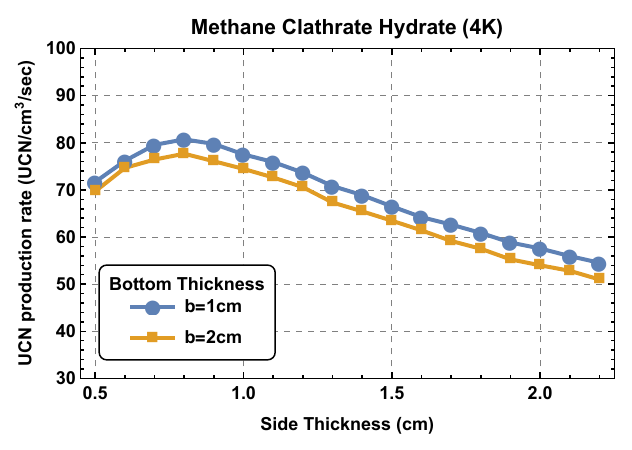}
} 
\caption
{The dependence of the UCN production rate on the cold moderator thickness in (a) solid $\mathrm{CH}_{4}$ cold moderator, (b) methane clathrate hydrate moderator, respectively. The blue circle is when the bottom thickness of the cold moderator is 1cm, and the orange square is when the bottom thickness of the cold moderator is 2cm. The side thickness of the cold moderator has been varied $0.5\mathrm{cm}\le s\le 2.2\mathrm{cm}$.}
\label{fig3} 
\end{figure}

\begin{figure}
\centering
\subfigure[Solid Methane (4K)]
{
    \label{fig7a}
    \includegraphics[width=7.5cm]{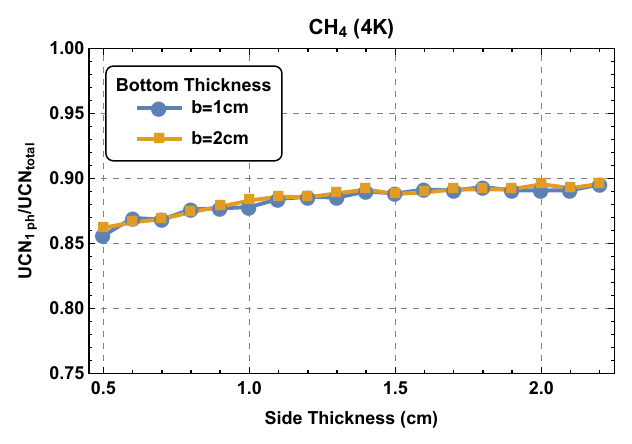}
} 
\hspace{1cm}
\subfigure[Methane Clathrate Hydrate (4K)] 
{
    \label{fig7b}
    \includegraphics[width=7.5cm]{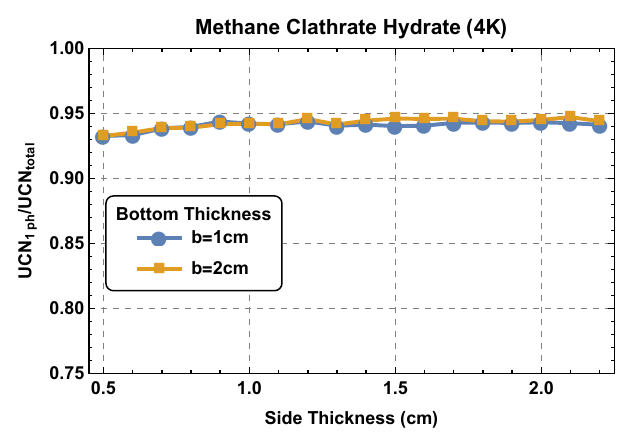}
} 
\caption
{The rate of UCN production from single-phonon scattering over the total UCN production in superfluid $^{4}\mathrm{He}$ UCN converter with (a) solid $\mathrm{CH}_{4}$ cold moderator, (b) methane clathrate hydrate moderator, respectively. The blue circle is when the bottom thickness of the cold moderator is 1cm, and the orange square is when the bottom thickness of the cold moderator is 2cm. The side thickness of the cold moderator has been varied $0.5\mathrm{cm}\le s\le 2.2\mathrm{cm}$.}
\label{fig7}
\end{figure}

Fig. \ref{fig3} shows the UCN production rate as a function of $b$ and $s$ for each moderator. In Fig. \ref{fig3a}, the UCN production rate from solid methane is as high as $\mathrm{P}_{\mathrm{UCN}}=54\mathrm{UCN}/\mathrm{cc}/\mathrm{s}$ when $b=1\mathrm{cm}$. But  the rate increases to $\mathrm{P}_{\mathrm{UCN}}=56\mathrm{UCN}/\mathrm{cc}/\mathrm{s}$ when $b=2\mathrm{cm}$. With methane clathrate hydrate cold moderator, the UCN production rate is $\mathrm{P}_{\mathrm{UCN}}=76\mathrm{UCN}/\mathrm{cc}/\mathrm{s}$ when $b=2\mathrm{cm}$. But, it increases to $\mathrm{P}_{\mathrm{UCN}}=80\mathrm{UCN}/\mathrm{cc}/\mathrm{s}$ when $b=1\mathrm{cm}$ as shown in Fig. \ref{fig3b}.

Fig. \ref{fig7} shows the ratio between the UCN production from the single-phonon excitation and the total UCN production, $\mathrm{UCN}_{1\mathrm{ph}}/\mathrm{UCN}_{\mathrm{tot}}$ for each moderator. When solid methane is used as the cold moderator, the contribution of the single-phonon excitation channel to UCN production varies from $85\%$ to $90\%$. However, this rate increases to $93\%\sim95\%$ when methane clathrate hydrate is used as the cold moderator. In both cases, the single-phonon excitation is the dominant process for UCN production over the multi-phonon process in the compact UCN source mainly because the cold neutron spectra produced from both cold moderators have a peak brightness near $1\mathrm{meV}$. 

After the UCNs are produced, the accumulation of UCNs for a long time in the superfluid $^{4}\rm{He}$ volume is in principle possible due to the zero nuclear absorption cross section of $^{4}\rm{He}$. If the upscattering of UCNs is sufficiently suppressed by lowering the temperature of $^{4}\rm{He}$ below $0.8\mathrm{K}$, the UCN storage time could be several hundred seconds. However, many previous experiments have shown that it is quite difficult to reach several hundred seconds of UCN storage time, mainly due to the collision between the UCNs with the inner surfaces of the storage volume. In recent experiments with a superfluid $^{4}\mathrm{He}$ UCN source in which the converter has a total length of about $70\mathrm{cm}$, an inner diameter of $66\mathrm{mm}$ and a Fermi potential of $184\mathrm{neV}$, a UCN storage time of $\tau=66\mathrm{sec}$ has been achieved at $0.8\mathrm{K}$\cite{Zimmer_2007,Schmidt-Wellenburg2009}.

Since the UCN converter volume has very similar dimension with one we suggest here, one can expect similar storage time in this  study as well. If we assume the UCN storage time $\tau$ as $\tau\approx50\mathrm{sec}$, the UCN density  becomes $\rho=\mathrm{P}\times\tau\approx\mathrm{P}_{\mathrm{UCN}}\times 50 \mathrm{sec}$. From the maximum UCN production rate, $\mathrm{P}_{\mathrm{UCN}}=80\mathrm{UCN}/\mathrm{cc}/\mathrm{s}$, the UCN density in the UCN converter is approximately $\rho_{\mathrm{converter}}=4000 \mathrm{UCN}/\mathrm{cc}$. Since we assumed a UCN production volume of $\sim1350\mathrm{cc}$, the total number of UCN produced in the UCN converter can be as high as $N_{\mathrm{converter}}\sim 5\times10^{6}\mathrm{UCN}$.

For the extraction of UCNs, a vertical extraction could be applied\cite{2010EPJC...67..589Z,2016PhRvC..93b5501L}. The vertical extraction method of UCNs from a converter was successfully demonstrated in recent experiments\cite{2010EPJC...67..589Z,2014PhRvC..90a5501P}. The main components necessary to realize vertical extraction are a short vertical section of the neutron guide located on top of the UCN converter and a cold valve that causes the UCNs to accumulate in the converter before they are extracted. Those components can be installed at the end of the UCN storage volume and operated with minimal temperature fluctuation. In the experiments of the vertical extraction, the $40\%$ of extraction efficiency $\epsilon$ was already achieved for UCNs accumulated in the storage volume at $0.8\mathrm{K}$\cite{2016PhRvC..93b5501L,2010EPJC...67..589Z}. With using the estimated efficiency $\epsilon=0.4$, the total number of UCNs at the end of the extraction line can be as high as $N_{\mathrm{extraction}}=\epsilon \times N_{\mathrm{converter}}\approx 2\times 10^{6} \mathrm{UCN}$ in the compact UCN source. 

%%%%%%%%%%
\section{\label{sec:heat}Radiation Heating}
%%%%%%%%%%
The heat loads on the UCN converter volume and the cold moderator volume in each setup were also studied. With regard to the heat load on the superfluid $^{4}\mathrm{He}$, the total heat load including the heat load from charged particles was estimated with MCNPX by tallying the energy deposition from the neutrons. 

\begin{figure}
\centering
\subfigure[Solid Methane (4K)] 
{
    \label{fig4a}
    \includegraphics[width=7.5cm]{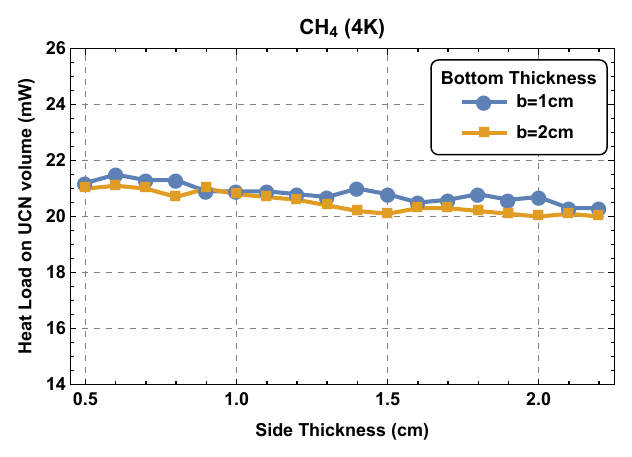}
} 
\hspace{1cm}
\subfigure[Methane Clathrate Hydrate (4K)] 
{
    \label{fig4b}
    \includegraphics[width=7.5cm]{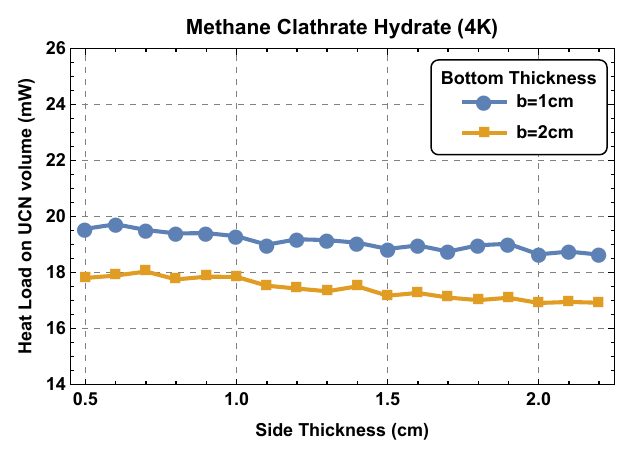}
} 
\caption
{Heat load on the superfluid $^{4}\mathrm{He}$ UCN converter with (a) solid $\mathrm{CH}_{4}$ cold moderator, (b) methane clathrate hydrate moderator, respectively. The blue circle is when the bottom thickness of the cold moderator is 1cm, and the orange square is when the bottom thickness of the cold moderator is 2cm. The side thickness of the cold moderator has been varied $0.5\mathrm{cm}\le s\le 2.2\mathrm{cm}$.}
\label{fig4} 
\end{figure}

Fig. \ref{fig4} shows the heat load on the UCN converter volume with solid methane and methane clathrate hydrate, respectively. Even with varying the side and bottom thicknesses of the cold  moderator volume, the overall heat load on the UCN converter volume remains less than 22mW in both cases. This extremely low heat load on the UCN converter volume would allow to operate the UCN source for longer period of time with reduced amounts of $^{3}\mathrm{He}$ circulation in evaporative $^{3}\mathrm{He}$ refrigeration.

\begin{figure}
\centering
\subfigure[Solid Methane (4K)]
{
    \label{fig5a}
    \includegraphics[width=7.5cm]{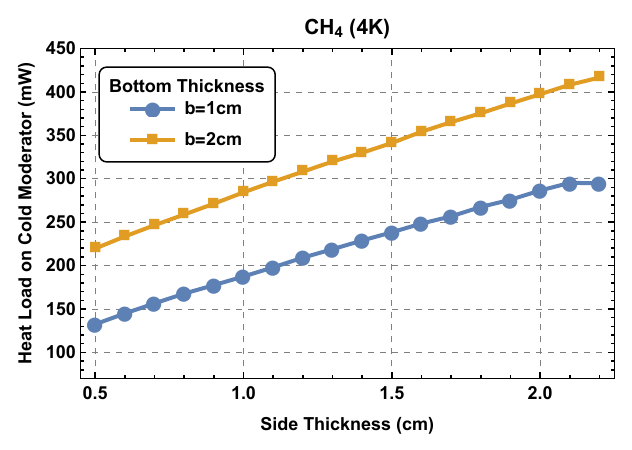}
} 
\hspace{1cm}
\subfigure[Methane Clathrate Hydrate (4K)]
{
    \label{fig5b}
    \includegraphics[width=7.5cm]{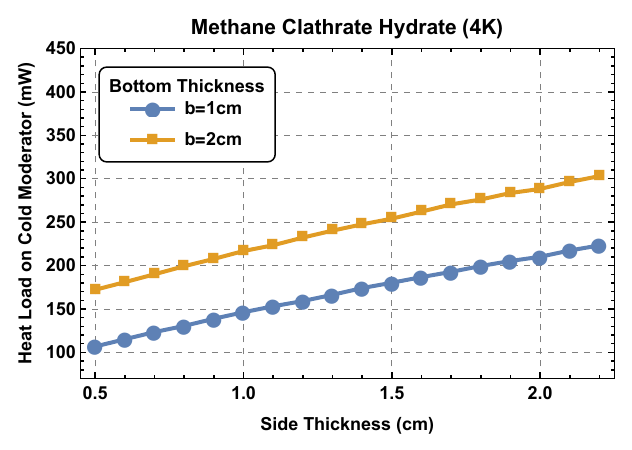}
} 
\caption
{Heat load on the cold moderator with (a) solid $\mathrm{CH}_{4}$ cold moderator, (b) methane clathrate hydrate moderator, respectively. The blue circle is when the bottom thickness of the cold moderator is 1cm, and the orange square is when the bottom thickness of the cold moderator is 2cm. The side thickness of the cold moderator has been varied $0.5\mathrm{cm}\le s\le 2.2\mathrm{cm}$.}
\label{fig5}
\end{figure}

The heat load on the cold moderator volume is also shown in Fig. \ref{fig5}. As side thickness becomes thicker, the overall heat load also increases to 450mW in the case of solid methane, and to 300mW in the case of methane clathrate hydrate. For the solid methane moderator, the thermal load at the optimal thickness for the maximum UCN production is as low as 350mW. In the case of methane clathrate hydrate, the thermal load drops to 200mW with the optimal thickness for maximum UCN production. Because of this very low thermal load on the cold moderator, the requirement for the refrigeration power of the cold moderator in the compact UCN source can be much less arduous than UCN sources cited at spallation or reactor facilities.

%%%%%%%%%%%%%%%
\section{\label{sec:con}Conclusion}
%%%%%%%%%%%%%%%
We proposed a new concept of UCN production in a compact UCN source. The compact UCN source includes a superfluid $^{4}\mathrm{He}$ UCN converter, and a layer of the cold moderator in the vicinity of low energy neutron production target with $13\mathrm{MeV}$ proton beam. The UCN production rate in the compact UCN source was estimated as high as 56 UCN/cc/sec with solid methane cold moderator. But the rate could get better to 80 UCN/cc/sec when methane clathrate hydrate was used as the supercold moderator medium. 

More than $85\%$ of UCNs were produced through single-phonon excitation in superfluid $^{4}\mathrm{He}$ in both cases. The thermal load on the UCN converter volume could be kept down below $22\mathrm{mW}$ in both cases. The heat load on the cold moderator also remains less than $400\mathrm{mW}$. This very low heat load on the UCN converter volume would enables the operation of the UCN source for an extended period of time. With using 40\% of UCN extraction efficiency and $50\mathrm{sec}$ of UCN storage time from previous experiments, the proposed compact UCN source could produce up to $\sim2\times10^{6}$ UCNs. Moreover, this compact UCN source would be still advantageous in terms of long-term operation as well as low operating cost. Our study is also of potential relevance for the laser-driven neutron sources envisioned in the future\cite{2018NIMPA.909..323C}.

%%%%%%%%%%%%%%%% 
\section{\label{sec: acknowledgement} Acknowledgement}
%%%%%%%%%%%%%%%% 
This work was supported by the Institute for Basic Science under Grant No. IBS-R017-D1-2021-a00. The work of W. Michael Snow is supported by NSF Grant No. PHY-1614545 and PHY-1914405 and by the IU Center for Spacetime Symmetries.

%%%%%%%%%%%%%%%% 

%%%%%%%%%%%%%%%% 
%%%%%%%%%%%%%%%% 
\end{document}